\documentclass[12pt]{article}
\usepackage{amsmath,amssymb,graphicx,mathrsfs,hyperref,slashed,color}
\newcommand{\be}{\begin{equation}}
\newcommand{\ee}{\end{equation}}
\newcommand{\bea}{\begin{eqnarray}}
\newcommand{\eea}{\end{eqnarray}}

\def\({\left(} \def\){\right)}

\renewcommand{\baselinestretch}{1.25}
\begin{document}
\title{\vspace{-1.8in}
{Discovering the interior of black holes}}
\author{\large Ram Brustein${}^{(1)}$,  A.J.M. Medved${}^{(2,3)}$, K. Yagi${}^{(4)}$
\\
\vspace{-.5in} \hspace{-1.5in} \vbox{
 \begin{flushleft}
  $^{\textrm{\normalsize
(1)\ Department of Physics, Ben-Gurion University,
    Beer-Sheva 84105, Israel}}$
$^{\textrm{\normalsize (2)\ Department of Physics \& Electronics, Rhodes University,
  Grahamstown 6140, South Africa}}$
$^{\textrm{\normalsize (3)\ National Institute for Theoretical Physics (NITheP), Western Cape 7602,
South Africa}}$
$^{\textrm{\normalsize (4)\ Department of Physics, Princeton University, Princeton, New Jersey 08544, USA}}$
\\ \small \hspace{1.07in}
    ramyb@bgu.ac.il,\  j.medved@ru.ac.za,\ kyagi@princeton.edu
\end{flushleft}
}}
\date{}
\maketitle

\begin{abstract}

The detection of gravitational waves (GWs) from black hole (BH)  mergers provides an inroad toward probing the interior of  astrophysical BHs. The general-relativistic description of the BH interior is that of  empty spacetime with a (possibly) singular core. Recently, however,  the hypothesis that the BH interior does not exist  has been gaining traction, as it provides a  means for resolving the BH information-loss problem. Here, we propose a simple method for answering the following question: Does the BH interior exist and, if so, does it contain some distribution of  matter or is it mostly empty? Our proposal is  premised on the idea that, similar to the case of relativistic, ultra-compact stars, any BH-like object whose interior has some matter distribution should support fluid modes  in addition to the conventional spacetime modes. In particular, the Coriolis-induced Rossby (r-) modes, whose spectrum is  mostly insensitive to the  composition of the interior matter, should be a universal feature of such BH-like objects. In fact, the frequency and damping time of these modes are determined by only the object's mass and speed of rotation. The r-modes oscillate at a lower frequency, decay at a slower rate and produce weaker GWs than do the spacetime modes. Hence, they imprint a model-insensitive signature of a non-empty interior in the GW spectrum resulting from  a BH merger. We find that future GW detectors, such as Advanced LIGO with its design sensitivity, have the potential of detecting such r-modes if the amount of GWs leaking out quantum mechanically from the interior of a BH-like object is sufficiently large.

\end{abstract}
\newpage
\renewcommand{\baselinestretch}{1.5}\normalsize

\subsection*{1. Introduction}

The view in  general relativity (GR) of a black hole (BH) as  a region of empty space except for a highly dense and classically singular core of matter has recently been presented with a formidable challenge --- it appears to be in contradiction with the laws of quantum mechanics!  The modern point of view  for diffusing this crisis is that the interior does not exist on account of spacetime ending at the BH horizon. There is, however, some divergence of views on how  spacetime  terminates.  Some argue that it ends with a ``firewall'' of high-energy particles surrounding the horizon \cite{AMPS} (also \cite{Sunny,Braun,MP}). Others argue that part of the geometry simply does not exist  as in the fuzzball model of BHs \cite{Mathur1,Mathur2,Mathur3,Mathur4} (also see \cite{otherfuzzball} and, more recently, \cite{hooft}).

But what if the BH interior does exist and is filled with some distribution of matter? The first obvious obstacle is how to prevent the inevitable fate of gravitational collapse that awaits any matter distribution whose size is approaching its gravitational radius \cite{buchdahl}.  What is then required is some exotic spacetime containing equally exotic matter which can be stored in an ultra-compact object that is able to withstand gravitational collapse. This object must, at the same time, exhibit all of the standard properties of BHs when viewed from the outside. We will refer to such spacetimes collectively as ``BH-like objects''. One  example for such an object is described  by our recent proposal that a BH should be modeled as a  bound and metastable state of highly energetic, interacting, long, closed strings; figuratively, a collapsed polymer \cite{strungout,emerge}.  (Here, we are using  ``collapsed'' as it is meant in the polymer literature, {\em e.g.} \cite{polymer}, and not gravitationally so.)

One can understand on a from a physics perspective  how such a stringy  object might evade gravitational collapse. A hot bath of closed strings  will entropically favor a state with just  a few long loops. These long strings can be effectively described as performing a random walk whose linear size, for a fixed total length of the strings, scales in four dimensions with the square root of the total length of the string.  In the case of the polymer model, this means that the linear size  of the region occupied by the strings  scales with the Schwarzschild radius. We are
then assuming that this  effective and repulsive random-walk ``force'' is enough to overcome the would-be gravitational collapse. We are also assuming that, like any other polymer,
a fluid-like description should be applicable, if only in a macroscopic, coarse-grained
sense.

Here the collapsed-polymer model is meant only as an illustrative example of a possibly more
general situation; namely, a BH proxy that is composed  of fluid-like matter.
It will eventually become clear that the analysis applies for this broader range
of models.

Putting such claims to the test need no longer be limited to the purview of  thought experiments and computer simulations. Thanks to the recent advancement in gravitational wave (GW) astronomy, brought to the forefront by the celebrated observation of GW150914 \cite{LIGO} and its companions \cite{LIGOII,Abbott:2016nmj}, there are  reasons to be optimistic about the prospects for future detections.  Indeed, the current observations have already proven their utility for constraining deviations from the GR  model of  BHs \cite{ligoGR,Pretorius}.

Let us briefly review as to why GWs can be  expected to carry information about their BH sources. (A longer discussion appears  in \cite{collision}.) The post-merger stage of a BH collision is that of a single  BH  settling down into a state of equilibrium. As is typical for partially open systems, the return to equilibrium is associated  with a set of   ringdown modes whose characteristic frequencies are determined by the system's  size, shape and composition.  These modes are necessarily damped and  often called quasinormal modes (QNMs). BHs are partially open in the sense that matter can enter but not exit  whereas, normally, the opposite is true. This makes BHs quite different from other partially open systems because the modes are not escaping from the BH itself, which is of course an impossibility. Rather, spacetime modes propagating in from infinity are reflected back from the surrounding  gravitational potential barrier or, otherwise,  transmitted through it. Whereas the transmitted modes continue on past  the horizon and are gone forever, some of the reflected modes constitute the observed GWs.

The frequencies and damping times for the reflected modes are determined by the properties of the gravitational potential barrier and, therefore, by only a handful  of BH parameters. Provided that the BH carries no net charges (nor any exotic ``hair''), the only relevant parameters are its mass $M$ and angular velocity   $\Omega$.  For further reading, one can start with the excellent review articles \cite{QNMBH,QNMBH2,Kony} and then, for example, \cite{Thorne,LD,KSmodel,Kojak,KSstar,LNS,inversecowling,Kokk1}.

The arguments in the current paper are premised on the idea that a BH-like object --- which is assumed to contain a non-trivial matter distribution rather than just a singular core --- has some resemblance  to a relativistic star.  As such, a BH-like object will have a collection of  fluid modes in addition to the previously described set of spacetime  modes, just like a relativistic star has both. In the relevant literature, the spacetime modes are called $w$-modes. As for the fluid modes, there are many different classes, with each class representing a different restoring force acting on  the star to  return it  to equilibrium. An incomplete list includes pressure ($p$-) modes, buoyancy or gravitational-restoring ($g$-) modes, shear ($s$-) modes and torsional ($t$-) modes. For most of these cases, the frequencies and damping times of the modes  are sensitive to the precise composition of the stellar  object.

Our current interest is  the spectra of the so-called $r$-modes ({\em e.g.}, \cite{PP,Saio,LS}). These are non-radial modes whose  amplitudes grow from zero at the center of the star to a maximal value at the surface. Their leading-order frequencies are  insensitive to the interior composition and, just like the spacetime modes of a BH,  depend only on the mass and rotational speed of the stellar body. So that, if one wants an answer to a  simple binary question --- ``does a BH-like object  contain a non-trivial matter distribution or does it not?''  --- these modes are just what is needed.

The $r$-modes are Rossby (planetary-like) waves that arise due to the effects of the Coriolis force; these being  the dominant effects of rotation provided that the object's radial velocity is smaller than the speed of light $c$.  This is because the Coriolis force is proportional to $\Omega$, whereas the effects of the centrifugal force are proportional to $\Omega^2$. As a consequence, a stellar body that is rotating slower than the speed of light can be treated, approximately,  as a spherically symmetric rotator. In a case where the axis of rotation points north, the Coriolis force induces counter-clockwise motion for fluid initially flowing to the north pole from the equator and  clockwise motion in the opposite case. One complete cycle defines the characteristic frequency of the mode, which scales linearly with $\Omega$.

One might wonder about the other types of fluid modes. These would
also be interesting for the purposes of discriminating between different models.
But, as other types of internal modes do depend on the composition of
the object, they would not have the same type of universality
that is being exploited here.
One might also wonder about the spacetime modes. But these, by definition,
do not know  about the details of the internal composition, as they depend strictly on the exterior geometry and boundary conditions at the outer surface.
The former is the same for any BH-like object, whereas the latter is a model-dependent consideration; for instance, some models are supposed  to produce ``echoes'' (see below). Nonetheless, a sufficiently compact object can be expected to produce modes that are similar to the predominant modes of a Kerr BH.

In the remainder of the paper, we review some basic facts about $r$-modes, both  in general and in the current context (Sec.~2), determine the characteristic properties of the resulting GWs (Sec.~3), present a gravitational waveform along with a plot of the associated  spectrum (Sec.~4),  discuss the prospects for detecting $r$-modes in the near future (Sec.~5) and then conclude (Sec.~6).

Before proceeding any further, it is important to emphasize that our
compact objects of interest are those whose outer surfaces act (at least effectively) as BH horizons in that they
inhibit the leakage of matter from inside to outside when
only the effects of general relativity are considered.  Our collapsed polymer
model has just such a ``quantum horizon''; its outer surface does not permit
matter to escape by classical means  but is otherwise only partially opaque
for finite $\hbar$ \cite{emerge}. This is because matter can only escape
as a result of string interactions, which is controlled by the string coupling,
a strictly quantum parameter. More generally,
a  quantum horizon refers to the  outer surface of a
BH-like object for which the escape of matter is a quantum process ---  quantum in the sense that it can be parametrized
by a small, dimensionless  parameter which  would not be present for the BHs of general relativity. We commonly refer to this small
 parameter
as a ``dimensionless $\hbar$'', which is simply the square of the string coupling for the collapsed-polymer model.
 (It is also assumed that
the fundamental spacetime modes of the object are close enough to those of a Kerr BH
so as to not yet be ruled out by the observational data.)   And it is this   quantum transparency
that will allow for the internal modes to couple to external GWs; albeit with
an appropriate suppression. This point is discussed further in Section~3,
although a full explanation will be  deferred until a later article
\cite{ridethewave}, where the same picture is considered from
the perspective of both an internal and  external observer.

Let us also  take note of a  different approach
\cite{Cardoso,Cardoso:2016oxy,Maggio:2017ivp} (also \cite{noncardoso,Abedi:2017isz})
which argues  that, for  ``exotic compact objects'' without horizons, there is a new class of modes that are absent in the classical-GR BH case and analogous to echoes ({\em i.e.}, modes trapped between the object's outer surface and potential barrier for a finite time). The basic idea is to model the interior of the object  as  a wormhole, as  the inner light ring of a wormhole captures the essence of an echo chamber.  Given this setup, one finds that the damping times of the trapped modes depend on a certain power of the log of the throat location
relative to  the Schwarzschild radius \cite{Pani-private}. Due to this large power, such a deviation in the damping times from the BH case effectively enters as a power-law deviation \cite{Maggio:2017ivp}. As shown in  a companion article  \cite{collision}, the collapsed-polymer model
also predicts power-law deviation to the damping times, albeit with a
much different expansion parameter.

\subsection*{2. The $r$-modes of a rotating black hole-like object}

A rotating BH-like object can be treated, approximately and to leading order in $\Omega$, as a spherically symmetric rotator with a constant angular velocity. Such a rotator naturally supports $r$-modes. Corrections to the leading order in $\Omega$ are expected to be of order $\Omega^2$. We will argue later that for the cases of interest, such corrections are small and therefore justify this approximation. Since our goal is to demonstrate how one could discriminate a fluid-filled interior from others in simple terms, we will confine ourselves to the non-relativistic approximation that allows us to obtain closed form expressions for the frequency and life-time of the $r$-modes. A more precise analysis may  be required for the purpose of making definitive quantitative predictions.

Closely following  \cite{Saio}, let us now review how these $r$-modes come about.

The starting point is the hydrodynamic momentum equation in the co-rotating frame of reference ({\em e.g.}, \cite{LL}). To leading order in the angular velocity
${\vec{\Omega}}$, this can be written as~\footnote{We are taking some liberty in using  a non-relativistic (Newtonian) equation to calculate the $r$-mode spectrum of BH-like objects. Our justification being that the production of $r$-modes  is, at leading order, a surface effect and thus insensitive
to what lies inside.  For reference, relativistic corrections only affect the
$r$-mode frequency of a neutron star by 8--20\% \cite{Idrisy:2014qca,Jasiulek:2016epr}, although these corrections would be enhanced for the case of a BH.
Ultimately, one would have to resort to the
numerical analysis of  the relativistic equations to make definitive predictions. Such a study is outside the scope of the current paper, which is meant
to convey the basic idea of using GWs to discriminate between
fluid-filled interiors and other models.}
\be
\frac{\partial {\vec u}}{\partial t}\;=\; -{\vec \nabla}{\delta\Phi}
 -\frac{1}{\rho}{\vec \nabla}{\delta p} +\frac{\delta \rho}{\rho^2}
{\vec \nabla} p
-2{\vec \Omega}\times {\vec u}\;,
\label{momentumhydro}
\ee
where  $\delta\Phi$ represents a  perturbation of the gravitational
potential, $p$ and $\delta p$ are the pressure and its  perturbation, $\rho$
and $\delta \rho$  are the energy density and its perturbation, and ${\vec u}$ is the velocity of the  fluid. It will be  assumed that  $\;\vec\Omega=\Omega\; \widehat{z}\;$ with
$\;\Omega>0\;$.

Let us now consider the radial component of the curl of Eq.~(\ref{momentumhydro}). With the approximations that non-radial motion dominates over radial motion, $u_r\ll u_\perp$, and that $\vec\nabla\cdot {\vec u}$ is at least linear order in $\Omega$ (in fact, it scales as $\Omega^3$ for the $r$-modes \cite{Saio}), the resulting equation is
$\;
\frac{\partial Z}{\partial t}=- 2\left({\vec u}_{\perp}\cdot
{\vec \nabla}_{\perp}\right){\vec \Omega}_{r}
\;$,
where $\;Z=\left({\vec \nabla}\times {\vec u}\right)_r\;$ is  the radial component of the vorticity and a subscript $\perp$ stands for the non-radial components of the
vector. We also used the fact that $p$ for the background  only acts radially.
Since $\vec\Omega$ does not depend on time explicitly,
\be
\frac{d}{dt}\left(Z+2{\Omega}_r\right)\;=\;0\;,
\label{vort}
\ee
to linear perturbative  order.
The quantity in the brackets is the radial component
of the vorticity in an inertial frame, and so  Eq.~(\ref{vort}) makes
it clear that this
component is conserved.

One can deduce from Eq.~(\ref{vort}) the nature of the induced oscillations. Working in the co-rotating frame, let us suppose that a fluid element starts out at the equator ($\theta=\pi/2$) where it is  moving north. Then, initially, $Z$ is a constant because ${\Omega}_r=\Omega\cos{\theta}=0\;$ and we choose $\;Z=0\;$ for simplicity. Now, as the fluid element proceeds  upwards, $\;{\Omega}_r\;$ increases because of the factor of $\cos{\theta}$. From Eq.~(\ref{vort}), it follows that a negative vorticity is generated, corresponding to a clockwise rotation  of the fluid element. The element then rotates in such  a way that  it eventually  returns to the equator and continues its motion downwards,
only to come back up  to the equator and so on. This type of motion is described in several nice movies \cite{r-mode-movies}.

A more formal approach allows one to deduce the actual relationship between the $r$-mode frequency and $\Omega$. As an  $r$-mode is toroidal at leading order, its  velocity vector in the co-rotating frame can be approximately
decomposed as   \cite{AS}
\be
{\vec u}\;\simeq\;i \omega r {K_{\ell m}}\left(0,\frac{1}{\sin{\theta}}\frac{\partial Y^m_{\ell}}{\partial\phi},
-\frac{\partial Y^m_{\ell}}{\partial\theta}\right)\;e^{i\omega t}\;,
\label{uhydro}
\ee
where $\ell$, $m$ are the angular-momentum quantum numbers, the $Y$'s are spherical harmonics and  $K_{\ell m}$ is some smooth function of $r$ which is not relevant to our purposes.
When substituting Eq.~(\ref{uhydro}) into Eq.~(\ref{vort}), one
finds that the  leading-order frequency of the $r$-modes in the co-rotating frame
is
$\;
\omega = \frac{2m\Omega}{\ell(\ell+1)}
\;$.
In an inertial frame, the frequency translates into
$\;
\omega = \Omega\left(-m + \frac{2m}{\ell(\ell+1)}\right)
\label{inertial-w}
\;$.

Our main interest is the case of $\;\ell=2\;$, $\;m=2\;$, for which
\be
\omega_{r-mode} \;=\; -\frac{4}{3}\Omega\;.
\label{wlm2}
\ee
The fact that the frequency is negative is significant and may, under some circumstances, result in  an instability which amplifies the $r$-modes \cite{instability}. This possibility will not be discussed any further and the negative sign will be left off.

To  determine the value of $\Omega$ for rotating BH-like objects, we may borrow some of the standard results for Kerr BHs  ({\em e.g.},  \cite{visser0706.0622}).  This is because, as far as their external properties are concerned, BH-like objects and the BHs of GR should --- by our previous assumptions and definition
for the compact objects of interest  ---  be similar and, in some cases like the collapsed-polymer model, indistinguishable. In what follows, $v$ is
the rotational speed of the object and $u$ indicates  the speed of a mode.

For spinning BHs, the frequency of rotation is parametrized by
the measure of spin  $\;a=2\;v/c\;$
($a$ is the dimensionless Kerr parameter),
\be
M \Omega\; =\; \frac{a}{2\left(1+\sqrt{1-a^2}\right)}\;.
\ee
Then, for the $r$-modes (with $\;\ell=m=2$),
\be
M \omega_{r-mode} \;= \;\frac{2}{3}\frac{a}{\left(1+\sqrt{1-a^2}\right)}\;.
\label{r-mode-anal}
\ee

In merger events for  which the masses of the two colliding BHs are
approximately equal and
their initial (total) spin is small compared to their angular momentum,
the final spin parameter is $\;a \approx 0.7\;$ and
depends weakly on the ratio of the masses (see  \cite{berti0703053} for
details). Also, as  reported in \cite{LIGOII}, this value of $a$ is approximately what was  measured in the  three recently detected events assuming
classical-GR BHs.  Then, with
this choice,
\be
M \Omega \;=\;  0.20\;
\label{Omega-num}
\ee
for the BH-like object and
\be
M \omega_{r-mode} \;=\;  0.27\;
\label{rmode-num}
\ee
for the frequency of the  $r$-modes with $\;\ell=m=2 \;$
({\em cf}, Eq.~(\ref{wlm2})).

For such cases, the relativistic corrections
due to the centrifugal force or to any additional relativistic corrections are governed by the small number
$\;
\frac{v^2}{c^2}=\frac{a^2}{4}= 0.12 \left(\frac{a}{0.7}\right)^2
$\;.
The velocity  of an  $r$-mode  is somewhat larger than the rotational velocity of the object but still quite non-relativistic,
$\;\frac{u_{r-mode}^2}{c^2}= \frac{\omega^2_{r-mode}}{\Omega^2}\frac{v^2}{c^2}=\frac{16}{9}\frac{v^2}{c^2}\simeq 0.22 \left(\frac{a}{0.7}\right)^2\;$. This means that the expected relativistic corrections are less than about $25\%$ of the non-relativistic values. At the level of accuracy of this paper, this is sufficient. To obtain more precise results one has to resort to better analytic and numerical analysis that will take into account also the relativistic corrections.

We now want to compare the frequency of the $r$-modes in Eq.~(\ref{rmode-num}) to that of  the slowest-oscillating spacetime modes $\omega_{st}$. The value of the latter  frequency for
the case of $\;a=0.7\;$ can be found  in ({\em e.g.}) Table II of
 \cite{0512160},
\be
M \omega_{st} =0.53\;.
\ee
It follows that the frequency of an   $r$-mode is about half that  of the lowest-frequency spacetime modes in the  $\;a=0.7\;$ case,
\be
\frac{\omega_{r-mode}}{\omega_{st}}\;\simeq\; 0.5\;,
\label{ratiorst}
\ee
up to a small (known) dependence on the ratio of the masses
of the colliding  BHs.

\subsection*{3. Frequency, decay time and amplitude of the emitted gravitational waves}

We would now like to determine the three quantities that characterize the additional emission of GWs  due to the $r$-modes: frequency, decay time and amplitude.  We find that the frequency, which is the most robust prediction, scales roughly as $\; \omega_{r-mode} \sim \omega_{st}\; v/c\;$. The decay time scales as $\;1/\tau_{r-mode}\sim (1/\tau_{st})(v/c)^2\;$ and is less robust. The amplitude scales as $\;h_{r-mode}\sim h_{st} (v/c)^3\;$ and is the least robust prediction. (Here, we have been using $\;u_{r-mode}\propto v$.) Each of the  three quantities will  be discussed in turn.

Let us  first recall what was found for the frequency.
For GWs that are sourced by  $r$-mode oscillations,
this is  given by Eq.~(\ref{r-mode-anal}) in general  and, for
values of the spin parameter close to $\;a=0.7\;$, by Eqs.~(\ref{rmode-num})
and~(\ref{ratiorst}).
In the latter case,  we recall that $\;M\omega_{r-mode}=0.27\;$ or, equivalently, $\;\omega_{r-mode}/\omega_{st}\simeq0.5\;$.
We will thus use the scaling relation
\begin{equation}
\label{eq:omega-r}
\omega_{r-mode} \;\sim\; \frac{u_{r-mode}}{c} \omega_{st}\;\simeq\; 0.5
\;\omega_{st}\; \left(\frac{a}{0.7}\right)\;.
\end{equation}
Although such a relation is based on only a single choice of $a$ (namely, $\;a=0.7$), it can be checked that Eq.~(\ref{eq:omega-r}) recovers
the correct  value of  $\omega_{r-mode}$ in Eq. \eqref{r-mode-anal} for the choice  of ({\em e.g.})  $\;a=0.5\;$  to within 5\% accuracy.

Let us next move on to the decay time. In general, the decay time  $\tau$ of a mode can be estimated by the ratio of
its dissipated energy $\frac{dE}{dt}$ to its total energy $E$, $\;1/\tau =\frac{1}{E}\frac{dE}{dt}\;$. The decay time of the $r$-modes and, therefore, of their corresponding GWs
is determined by the shortest dissipation time of three possibly important sources of dissipation: (i) the emission of GWs which reduces the energy of the $r$-modes accordingly, (ii) the leakage of  $r$-modes  away from the BH-like object
by processes that differ from the emission of GWs (for instance,
by coupling to other types of matter) and (iii) the intrinsic dissipation within the interior matter.

First, the decay time of the $r$-modes  due to emission of GWs scales as the light-crossing time $R/c$ divided by  a factor of $(M\omega_{r-mode})^6$ ({\em e.g.}, \cite{instability}).
This is much too long a time scale to be of  any  relevance to our discussion.

Second, the time scale for leakage can be estimated by calculating the imaginary part of the QNM frequencies.
As explained in detail in \cite{collision}, when the modes are non-relativistic, the imaginary part of the frequency $\omega_I$ is parametrically smaller than the real part $\omega$ because of the  scaling
$\;
\omega_I \sim \frac{u}{c}\; \omega
\;$.
Then it follows from Eq. \eqref{eq:omega-r} that the imaginary part of the $r$-mode frequency is doubly suppressed relative to that of the spacetime QNMs,
\be
\omega_{I\;r-mode}\; \sim\; \left(\frac{u_{r-mode}}{c}\right)^{2} \omega_{I\;st}\;,
\ee
where the value of $\omega_{I\;st}$ for $a=0.7$ is given in ({\em e.g.}) Table II of \cite{0512160}, $\;M \omega_{I\;st}= 0.08 \;$.
Equivalently,
\be
\label{eq:tau-r}
\tau_{r-mode}\;\sim\; \left(\frac{u_{r-mode}}{c}\right)^{-2} \tau_{st}\;\simeq\; 4.6
\;\tau_{st}\; \left(\frac{a}{0.7}\right)^{-2}\;.
\ee

The third  source of energy loss is the intrinsic dissipation, whose time
scale  can be estimated following \cite{instability}. As will be shown, unless the ratio of the
 shear viscosity $\eta$ to the  entropy density $s$ is the smallest that it can be  --- an approximate  saturation of the KSS bound
$\;\eta/s\sim 1\;$ \cite{KSS}  ---  then the intrinsic dissipation is too large and it is likely that the modes will decay too quickly to ever be detected.
 In the case of the polymer model, the interior matter does indeed saturate the KSS bound \cite{emerge}, and
a simple argument (based on reinterpreting the KSS bound as an upper limit
on the entropy \cite{BekKSS}) suggests that this must be generally true
for other models as well.
Following  \cite{instability},  one then finds that the intrinsic-dissipation time
$\;{\widetilde \tau}\;$ for the $r$-modes is given by
\be
\frac{1}{{\widetilde \tau}_{r-mode}}\; \sim \; \frac{\eta}{\rho R^2}\;\sim\; \left(\frac{u_{r-mode}}{c}\right)^2 \frac{1}{\tau_{st}}\;,
\ee
where we have used  the fact that $\;\eta/\rho\sim R\;$ for KSS-saturating matter with relativistic modes
and  $\eta/\rho$  effectively scales like $\left(\frac{u}{c}\right)^{2}\;$ for non-relativistic modes \cite{collision} so that  $\;\eta/\rho\sim \left(\frac{u}{c}\right)^{2} R\;$.  If $\eta/\rho$ is parametrically larger than $R$, as is the case for all known forms of  non-exotic matter, then the  decay time would be much smaller than that  of the longest-lived spacetime modes, meaning that  the detection
of the $r$-modes would no longer be feasible.

Conversely, if  $\; \eta/\rho\sim R\;$ as expected, then both the leakage time and the intrinsic-dissipation time are  parametrically longer than the decay time of the spacetime QNMs  by a   factor of $(u/c)^{-2}$,
\be
\tau_{r-mode}\;\sim\; {\widetilde \tau}_{r-mode}\;\sim\; \left(\frac{u_{r-mode}}{c}\right)^{-2}\tau_{st} \; \simeq \; 4.6 \;\tau_{st}\; \left(\frac{a}{0.7}\right)^{-2}\;.
\ee

Let us now consider the amplitude of the emitted GWs. Our approach is
to use Einstein's celebrated quadrupole formula, while taking into account that the matter in some models can be surrounded by a (possibly semi-transparent) horizon. The latter consideration
can be incorporated by  parametrizing   the strength of the coupling of the fluid  modes to the emitted GWs. For any specific case, this coupling is determined by the details of the model. For example, if the matter within  a BH-like object is not surrounded by any horizon, this coupling can be  estimated by treating the background spacetime as fixed and (essentially) flat \cite{ThorneRMP}. Then, $\;h\propto d^2Q/dt^2\;$, where $h$ is the gravitational waveform and $Q$ is the quadrupole moment of the energy density.

Now suppose that some quadrupole moment does exist in a confined region of space. Just how much of this moment  contributes to the production of outgoing GWs? If the region is surrounded by a classical horizon, the answer is none. In this case, the horizon is completely opaque and nothing can escape from inside. On the other hand, the region will be semi-transparent if surrounded by a ``quantum horizon" because then some GWs can  escape to the outside. The fraction of those escaping is  proportional to the dimensionless $\hbar$ of the problem, $\;{\widetilde\hbar}< 1\;$.  For example, in the polymer model, the relevant dimensionless parameter for a certain class of fluid modes  is $\; {\widetilde \hbar}=g_s^2\;$ \cite{collision}, where the string coupling $g_s$ is the ratio between the Planck length and the string length scale. The numerical value of $g_s$ is expected to be small but not extremely small. For instance, the string coupling cannot be too much smaller than  unity given that the expected grand unification of forces at the Planck energy is correct.
In cases like the wormhole model \cite{Cardoso}, the region is not surrounded by any horizon.

We will  cover this broad spectrum of cases
by introducing  a  ``transparency" or transmission coefficient $T_{hor}$ that ranges
from  0 (a classical horizon) to 1 (no horizon). An $r$-mode can
now be characterized  as follows:  Its frequency and  lifetime are fixed by the frequency of rotation (equivalently, the Kerr parameter $a$) of the BH-like object,  whereas its  amplitude additionally depends on a  model-dependent parameter
$T_{hor}$  for which  $\;0\leq T_{hor}\leq 1\;$.

Let us briefly comment on how $T_{hor}$ can generally be estimated (see \cite{ridethewave} for a detailed discussion). One can assign a width to a given quantum horizon of $\;\Delta R_S= {\widetilde\hbar} R_S\;$ ($R_S$ is the object's Schwarzschild radius). The width $\Delta R_S$ can, when the BH is out of equilibrium,  be expected to be macroscopically large and still well within the potential barrier at about $\frac{3}{2}R_S$. This is because $\;\Delta R_S$  scales with the product of the  horizon radius and  a simple, positive  power of the  dimensionless $\hbar$ which need only be smaller than unity.    The width $\Delta R_S$  implies that the GWs corresponding to some fluid mode will first appear in the exterior at a radius where the   Tolman redshift factor is $\sqrt{\widetilde\hbar}$. Using this redshift along with the quadrupole formula, one finds that the amplitude of the GWs, by the time they reach the potential  barrier, will be suppressed by some power of  ${\widetilde\hbar}$ --- it is this suppression that should be identified with $T_{hor}$, a number that is  less than one but, at the same time, need not be unobservably small.  On the other hand, in  cases like the wormhole model for which there is no horizon, one can view $T_{hor}$ as some power of the redshift factor at the location of the object's outermost  surface or its throat.

The redshift factor describes how an  external observer, who believes that the fluid modes originate from  outside of the horizon, is able to reconcile the suppression factor $T_{hor}$ with her knowledge of general relativity. From an internal perspective, the suppression can be attributed to quantum  effects being the primary source of mode leakage. One should not combine these two sources of suppression, as this would amount to a  double counting. The consistency between the internal and external perspectives and that these provide complementary pictures will be
 exposed in the aforementioned treatment \cite{ridethewave}.

Putting all of these ingredients together and recognizing that the $r$-modes induce velocity perturbations,
one can find  an  appropriate estimate of the GW amplitude in \cite{ThorneRMP} (also \cite{instability}). Let us first express the $r$-mode waveform as
\be
h_{r-mode}\; =\; A_{r-mode}\, e^{-t/\tau_{r-mode}}\, \sin(\omega_{r-mode}\, t - \phi_r)\;,
\ee
with $\phi_r$ representing the constant phase and
 $A_{r-mode}$, the dimensionless strain amplitude. Then
\be
A_{r-mode} \;\sim\; \alpha_{r-mode}\ T_{hor} \frac{M}{r_s} \left(\frac{u_{r-mode}}{c}\right)^3 \;,
\ee
where  $\;\alpha_{I} < 1\;$ parametrizes the amount of energy that the merger
injects into the $I^{\rm th}$ class of mode perturbations and $r_s$  is the radial distance from the center of the  source. The factor $\left(\frac{u_{r-mode}}{c}\right)^3$ is a product of a factor of $\left(\frac{u_{r-mode}}{c}\right)^2$ originating from the two time derivatives in the quadrupole formula ($d/dt \sim \omega_{r-mode}\propto u_{r-mode\;}$) and additional factor of $\frac{u_{r-mode}}{c}$ that can be attributed to the waves being sourced by velocity perturbations.

This amplitude should be compared to that of the spacetime modes, which scales as
\be
A_{st} \;\sim\; \alpha_{st} \frac{M}{r_s}\;.
\ee
In the recently detected events, the fraction of radiant energy  in the form of  GWs was found to be a few percent of the system's total mass, which  is consistent with prior estimates of about $\;\alpha_{st}\sim 0.1\;$ corresponding to a gravitational radiant energy of around $3\%$ of $M$  \cite{lrrSchutz,FlannaganHughes}. It is likely that $\alpha_{r-mode}$ and $\alpha_{st}$ are of similar magnitudes; in which case, the suppression of the $r$-mode amplitude
is determined solely  by  $T_{hor} (u_{r-mode}/c)^3$,
\be
\label{eq:hr}
A_{r-mode}\;\sim\;T_{hor} \left(\frac{u_{r-mode}}{c}\right)^3 A_{st} \;\sim\; 0.1\; T_{hor}\; A_{st}\; \left(\frac{a}{0.7}\right)^3\;.
\ee

\subsection*{4. Gravitational waveform and spectrum}

Let us now look at the gravitational waveform for the $r$-modes in both the time and Fourier domain, beginning with the former.
The case of primary  interest is when the  final spin is $a=0.7$,  which corresponds to the merger of two non-spinning, equal-mass BHs.
 From the results of the previous
section, the following picture emerges: In a BH-merger event, the $r$-modes produce a GW signal at a lower frequency,
 $\;\omega_{r-mode} \sim 0.5\; \omega_{st}\;$,  with a longer decay time, $\;\tau_{r-mode} \simeq 4.6\; \tau_{st}\;$, and with a suppressed  amplitude, $\;h_{r-mode} \sim 0.1\; h_{st}\;$, in  comparison to the standard  spacetime-mode signal. One can also anticipate some additional  delay in the emission of GWs due to the reduction in  frequency, as there is an expected delay of about  one oscillatory period. (This allows time for the mode to reach the outer surface.)
Figure~1 depicts the  waveform of GWs emitted from a BH-merger --- if the $r$-modes do exist ---  for a final spin of $\;a=0.7\;$, $\;v/c=0.35\;$, $\;\omega_{r-mode} = 0.5 \;\omega_{st}\;$, $\;\tau_{r-mode} = 5 \;\tau_{st}\;$  and $\;h_{r-mode} = 0.1 \; h_{st}\;$, along  with a delay of about one  period.
\begin{figure*}[!ht]
\begin{center}
\includegraphics[width=.75\textwidth]{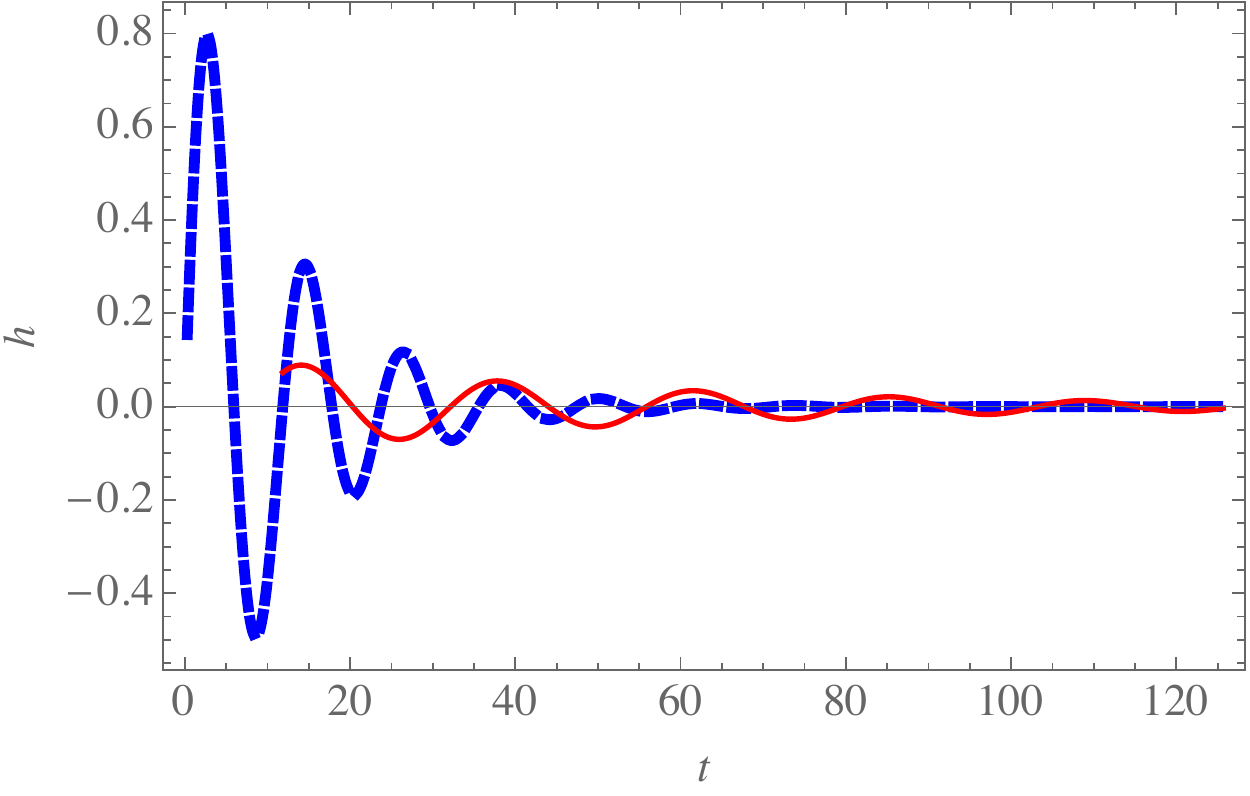}
\caption{Gravitational waves emitted during the ringdown phase of a BH merger with the parameters listed in the text. The blue (thick, dashed) line depicts  $h_{st}$ in arbitrary units as a function of time in units of $M$, while the red (thin, solid) line depicts $h_{r-mode}$.}
\end{center}
\end{figure*}

We next consider the GW spectrum in the Fourier domain, as this is important for calculating the signal-to-noise ratio (SNR) later. The Fourier transform of a damped sinusoid is given by \cite{Berti:2007zu,collision}
\bea
\label{eq:hf}
|\tilde h(f)| &=& A_{r-mode}\, \tau_{r-mode} \bigg| \frac{2f_r^2 Q_r \cos \phi_r- f_r (f_r-2 i f Q_r) \sin \phi_r}{f_r^2-4i f f_r Q_r + 4 (f_r^2-f^2)Q_r^2} \bigg| \;,
\eea
where $\;f_r \equiv \omega_{r-mode}/(2\pi)\;$ is the $r$-mode frequency and $\;Q_r \equiv \pi f_r \tau_{r-mode}\;$. To make this transform explicit, Eqs.~\eqref{eq:omega-r} and~\eqref{eq:tau-r} can be used to determine how the $r$-mode frequency and damping time scale
with respect to those of  the spacetime mode. Meanwhile,  the fitting function for the spacetime-mode parameters  in terms of $M$ and $a$ can be found in \cite{0512160}. Similarly, Eq.~\eqref{eq:hr} can be used for the amplitude scaling, where the spacetime-mode amplitude $A_{st}$ can be obtained from \cite{collision}.

\begin{figure}[!ht]
\begin{center}
\includegraphics[width=.75\textwidth]{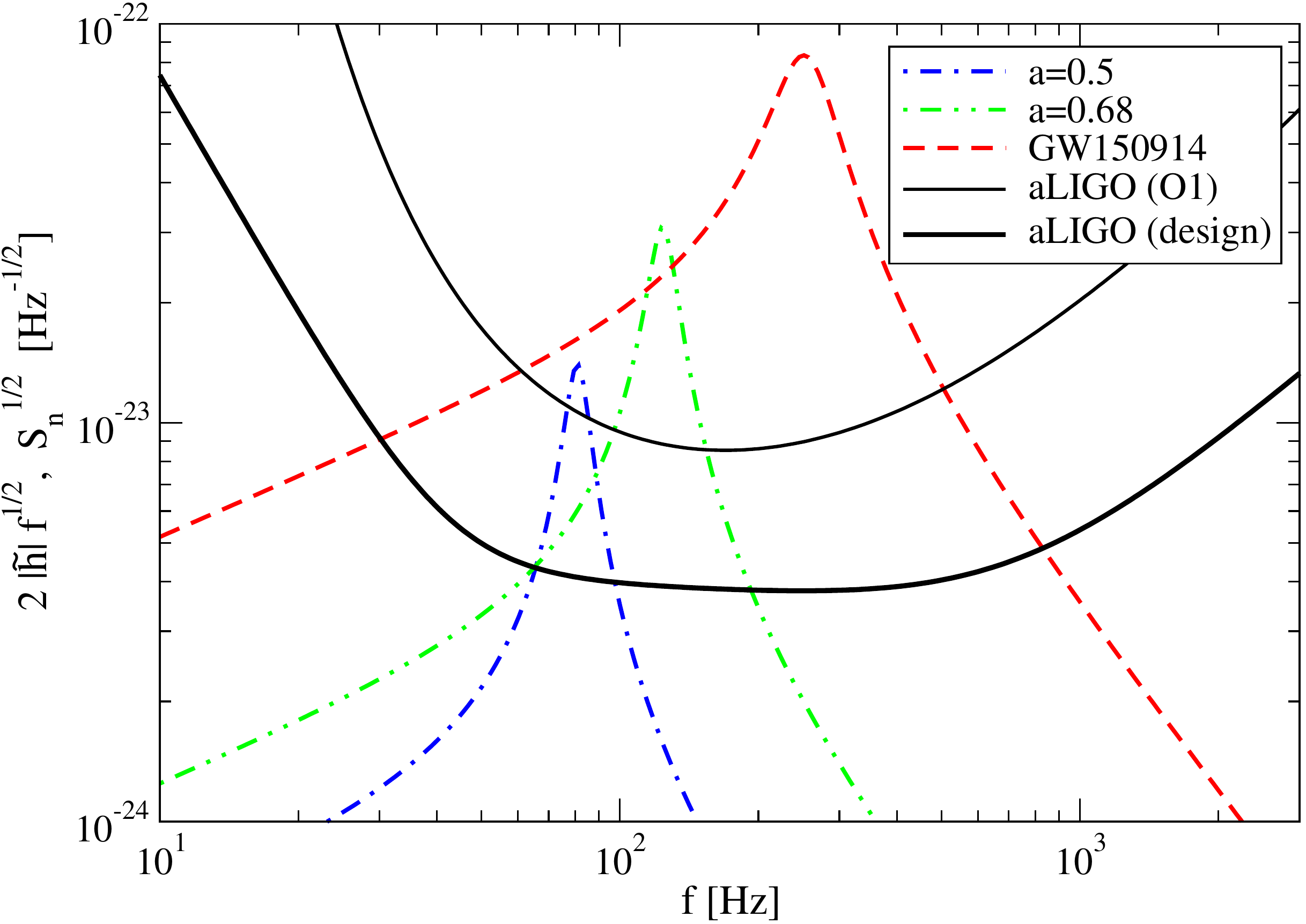}
\caption{\label{fig:spectrum}
Sky-averaged GW spectrum of the $r$-mode for BH-like objects with $a=0.5$ (blue, dotted-dashed) and $a=0.68$ (green, double dotted-dashed). We choose $M=62.3\;M_\odot$, $D_L=410$~Mpc and $T_{hor}=1$. For reference, the spectrum corresponding to the observed ringdown for GW150914 (red, dashed) is included. Also shown are  the noise spectral density of aLIGO in the O1 run (thin, black, solid) and for its design sensitivity (thick, black, solid). The ratio between the signal and noise roughly corresponds to the SNR and the signal is detectable if this ratio is above the threshold $(\sim 5)$. We stress that the results presented here are not robust and should be understood as only rough estimates.}
\end{center}
\end{figure}

Figure~\ref{fig:spectrum} presents such spectra for $\;a=0.5\;$ and $\;a=0.68\;$.
Here, we have set  $\;T_{hor}=1\;$, $\;\phi_r=0\;$,
depicted the sky-averaged amplitude at a luminosity distance of $\;D_L=410$~Mpc and chosen $\;M=62.3\;M_\odot\;$, where the last two values correspond to those of GW150914 \cite{LIGO,LIGOII}.
The relation between $a$ and the symmetric mass ratio $\eta$ of a BH binary \cite{Husa:2015iqa}  has been adopted to rewrite the radiation efficiency in $A_{st}$ (with the pre-merger BH spins set to 0 for simplicity) in terms of $a$. One can  observe how the amplitude, frequency and the width of the peak all grow with increasing
$a$. For reference, we have included the spectrum of the spacetime mode for GW150914; as well as  the noise spectral density  of  Advanced LIGO (aLIGO), both for its  O1 run and for  its design sensitivity.

\subsection*{5. Prospects for detection}

Let us now discuss the future prospects for detecting $r$-modes.
In \cite{collision}, we derive an upper bound on the amplitude of the secondary ringdown mode relative to the primary one assuming that the former was not detected in the GW150914 observation. Applying  that result to the current analysis
and  choosing  $a=0.68$ (the final spin of the remnant BH for GW150914 \cite{LIGO,LIGOII}), we then obtain $\;h_{r-mode}/h_{st} < 0.26\;$. This inequality can,  using Eq.~\eqref{eq:hr}, be mapped to one on $T_{hor}$, leading to $\;T_{hor} \lesssim 2.6\;$. This should be regarded as only a rough bound, as it is based on scaling relations for the amplitude, frequency and damping time which  neglect any $O(1)$ prefactors.
Rough or otherwise,  such a bound is not really useful because  $T_{hor}$
cannot be any larger than unity.

Our main interest is in the scaling relation for the minimum $T_{hor}$ that is  required for detecting  $r$-modes (which we denote  $T_{hor}^{(\min)}$)
in terms of $M$, $a$, $D_L$ and detector sensitivity.
The starting point is the calculation of
the SNR, which is obtained from
\be
\label{eq:SNR}
\mathrm{SNR}^2  \;=\; 4 \int_{f_\mathrm{min}}^{f_\mathrm{max}} \frac{|\tilde h (f)|^2}{S_n (f)} df\;,
\ee
where  $f_\mathrm{min}$ and  $f_\mathrm{max}$ are the minimum and maximum
frequencies, while $S_n$ is the noise spectral density.
Then using Eqs.~\eqref{eq:hr}--\eqref{eq:SNR}, along with  $\;u_{r-mode}/c \propto a\;$ and $\;df \sim 1/\tau_{r-mode}\;$, one finds that
\be
\label{eq:SNR-scaling}
\mathrm{SNR} \;\propto\; A_{r-mode} \sqrt{\tau_{r-mode}} \;\propto\; T_{hor}\, A_{st}\, a^3\, \sqrt{\tau_{r-mode}}
\ee
for a white-noise background. It is worth noting that the SNR scales as $\;(v/c)^2\;$ since $\;A_{r-mode} \propto (v/c)^3\;$ and $\;\tau_{r-mode} \propto (v/c)^{-2}\;$. One can now derive the minimum $T_{hor}$ for detection by equating this calculation to the threshold  SNR. However, our main interest  is still in the scaling behavior of $T_{hor}^{(\min)}$. For instance,  since $D_L$ only appears in Eq.~(\ref{eq:SNR-scaling}) through  $A_{st}$ as  $\;A_{st} \propto 1/D_L\;$, $T_{hor}^{(\min)}$ is linearly proportional to $D_L$.

\begin{figure}[!ht]
\begin{center}
\includegraphics[width=.75\textwidth]{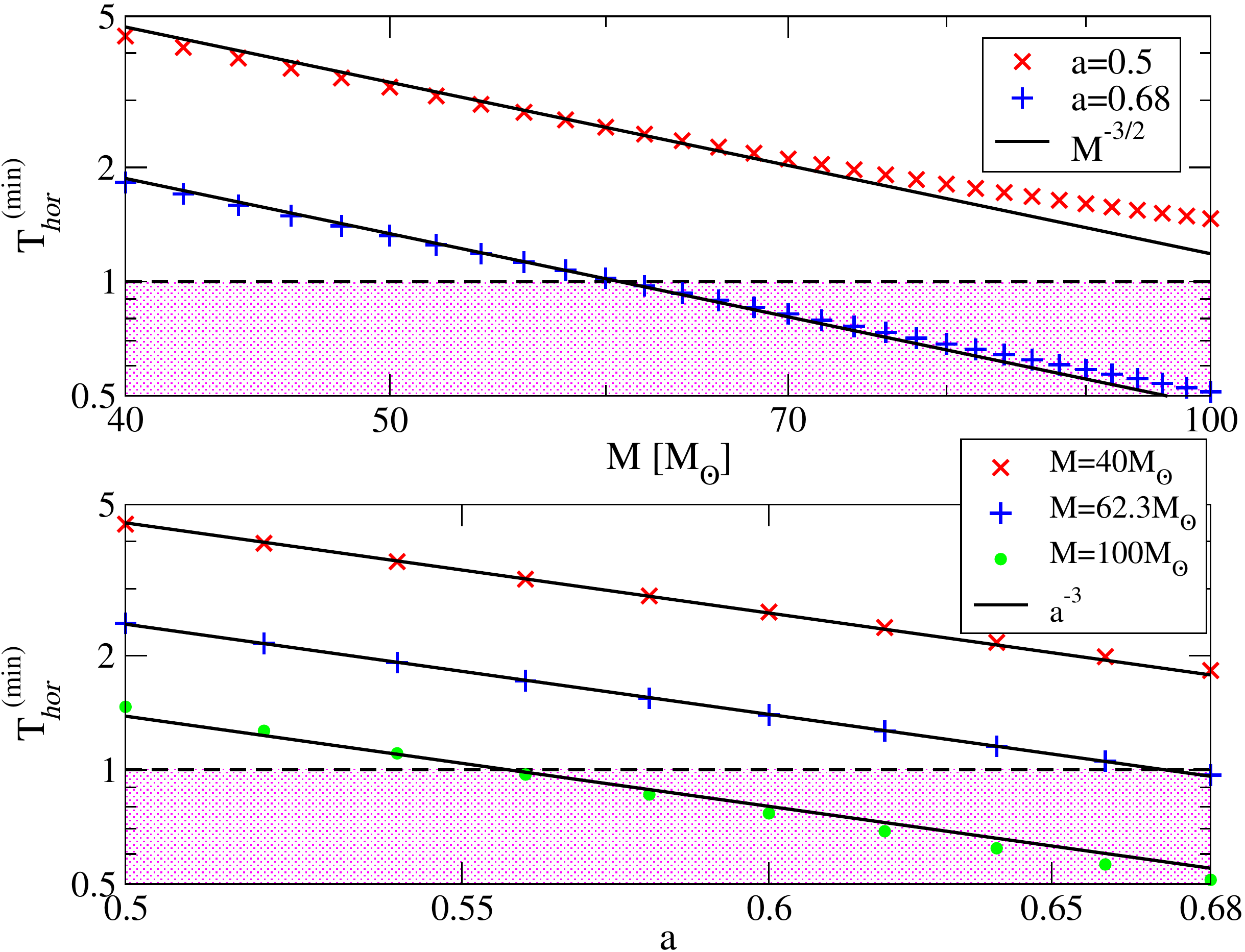}
\caption{\label{fig:M-a-dep}
The minimum $T_{hor}$, which characterizes the amount of quantum leakage of the $r$-mode GWs through the horizon, that is needed for aLIGO at Hanford and Livingston with its design sensitivity to detect $r$-mode GWs from equal-mass BH binaries at $D_L=410\;$Mpc. We show the minimum $T_{hor}$ as a function of $M$ (top) and $a$ (bottom). Solid lines are the fits proportional to $M^{-3/2}$ and $a^{-3}$. The shaded region $(T_{hor} \leq 1)$ corresponds to the theoretically allowed range of $T_{hor}$. Observe that $T_{hor}^{(\min)}$ for massive and rapidly spinning sources falls into this range. We stress that the bounds presented here are not robust and should be understood as only rough estimates.}
\end{center}
\end{figure}

Let us  next look at the $M$ dependence of $T_{hor}^{(\min)}$. Recalling that $\;A_{st} \propto M\;$ and $\;\tau_{r-mode} \propto M\;$, one can see from Eq.~\eqref{eq:SNR-scaling} that $\mathrm{SNR} \propto M^{3/2}$. Thus, setting this expression for the SNR equal to the threshold SNR of 5, one can deduce that $T_{hor}^{(\min)}$ is  proportional to $M^{-3/2}$. The top panel of Fig.~\ref{fig:M-a-dep} shows, for a fixed set of $a$ values, the $M$ dependence of $T_{hor}^{(\min)}$ as calculated {\em directly} from Eq.~\eqref{eq:SNR} ({\em i.e.}, without imposing the white-noise assumption or using Eq.~\eqref{eq:SNR-scaling}) for  a sky-averaged waveform. One can compare
this figure to the fit proportional to $M^{-3/2}$ (which is
also plotted) and observe that the numerical values follow the anticipated  $M^{-3/2}$ dependence for the smaller values of $M$. For larger $M$, the peak frequency of the GW spectrum in Fig.~\ref{fig:spectrum} shifts to a lower frequency and, as a result, the white-noise assumption becomes less valid. Thus, the minimum $T_{hor}$ for detection deviates from its expected $M^{-3/2}$ dependence in
this regime of larger mass.

Finally, we can consider the $a$ dependence of $T_{hor}^{(\min)}$.
For one thing,  $A_{st}$ is proportional to the radiation efficiency, which is further proportional to the symmetric mass ratio $\eta$ \cite{berti0703053},
which is roughly proportional to $a$ \cite{Husa:2015iqa}. For another,  $\;\tau_{r-mode} \propto a^{-2}\;$, and hence  Eq.~\eqref{eq:SNR-scaling} indicates that  $\;\mathrm{SNR} \propto a^{3}\;$. It then follows, in analogy to  the $M$-dependence
argument, that $T_{hor}^{(\min)}$ is proportional
to $a^{-3}$. The bottom panel of Fig.~\ref{fig:M-a-dep} depicts how $T_{hor}^{(\min)}$  depends on $a$ as calculated from Eq.~\eqref{eq:SNR} for a set of fixed $M$ values.  Also shown is  the fit proportional to $a^{-3}$. Once again,
the numerical values nicely follow  the anticipated  dependence  when $M$ is smaller but deviate from expectations when $M$ is larger.
The logic underlying this behavior  is, of course, the same as that discussed in the previous paragraph.

In light of its dependence on $M$, $a$ and $D_L$,  one can roughly estimate the minimum $T_{hor}$ for detection as
\begin{eqnarray}
\label{eq:Thor-min}
T_{hor}^{(\min)} \; &\approx& \; 0.97\; \left( \frac{M}{62.3 M_\odot} \right)^{-3/2} \left( \frac{a}{0.68} \right)^{-3}  \left( \frac{D_L}{410 \mathrm{Mpc}} \right) \nonumber \\
& &\times \left( \frac{N_s}{1} \right)^{-1/2}  \left( \frac{N_d}{2} \right)^{-1/2} \left( \frac{\sqrt{S_n(f_0)}}{4\times 10^{-24}\mathrm{Hz}^{-1/2}} \right)\;,
\end{eqnarray}
where we also included the dependence on the number of (identical) GW sources $N_s$ and the number of (identical) GW detectors $N_d$. See, for instance, \cite{Yang:2017zxs} on how to coherently stack small-SNR signals from different GW sources.  Additionally, $\sqrt{S_n(f_0)}$ is the detector sensitivity at $f_0=200\;$Hz and is merely a representative parameter for an overall sensitivity scaling (as $T_{hor}^{(\min)}$ depends on $\sqrt{S_n(f)}$ and not just $\sqrt{S_n(f_0)}$~).

Let us study the prospect for the detection of $r$-modes in more detail.
Equation~\eqref{eq:Thor-min} and Fig.~\ref{fig:M-a-dep} imply that the detectability increases for sufficiently massive, rapidly spinning and  close-enough objects. For such sources, $T_{hor}^{(\min)}$ becomes smaller than unity and falls into the theoretically allowed range of $T_{hor}$, as indicated by the magenta shaded regions in Fig.~\ref{fig:M-a-dep}. For example, a mass of $\;M=100~M_\odot\;$ allows one to detect an $r$-mode with $T_{hor}$ as small as $\sim 0.5$. If Virgo, KAGRA and LIGO-India further come online ($N_d=5$), an $r$-mode can be detected with $\;T_{hor} \gtrsim 0.3\;$. On the other hand, third-generation GW detectors, such as the  Einstein Telescope and Cosmic Explorer, will have $\sim 10$ times better sensitivity than aLIGO. Hence  an $r$-mode can be detected with $T_{hor}$ as small as $\sim 0.1$ for the fiducial parameters in Eq.~\eqref{eq:Thor-min} when using third-generation detectors. Alternatively, such detectors may find $\sim 10^3$ GW sources having  a similar SNR to that of GW150914 $(\sim 20)$. Setting $D_L$ ($\sqrt{S_n(f_0)}$~) to be  10 times larger (smaller) and $\;N_s=10^3\;$ in Eq.~\eqref{eq:Thor-min}, one finds that  $r$-modes can be detected with $\;T_{hor} \gtrsim 0.03\;$.

\subsection*{6. Conclusion}

We have argued that a BH-like object --- an object that resembles a BH from the outside but with a different composition for its  interior --- can be discriminated from the BHs of GR on the basis  of its  $r$-modes. This follows from the observation that, just like a relativistic star, the $r$-mode frequency and damping time
should be  essentially independent of the object's composition, depending only on its  mass and speed of rotation $v$ to leading order in $v/c$.

 Under suitable circumstances, the GWs originating from the $r$-modes should stand out clearly in the data, as their frequencies scale with the rotational speed of the BH-like object and their lifetimes are enhanced by a factor of $(v/c)^{-2}$.  However, because the wave amplitude drops off quickly by a factor of $(v/c)^{3}$, one is faced with two competing effects: The easier it is to distinguish the $r$-mode-sourced GWs  from those sourced by the spacetime modes, the weaker is the $r$-mode signal. The GW spectrum also drops out of the detector band for smaller $v/c$, making the detection of such  lower-frequency waves even more difficult. More optimistically, we have shown that, given aLIGO's design sensitivity and  a sufficiently massive, rapidly rotating and close-enough source, the minimum value of $T_{hor}$ --- this being
a parameter which characterizes the quantum leakage of the $r$-mode GWs through the horizon --- that is needed for detection is below unity, which is the theoretical upper bound on $T_{hor}$. The prospect for detection increases as the detector sensitivity improves, more detectors come online and the number of GW sources increases. Alternatively, the absence of any $r$-modes would allow  one to place upper bounds on $T_{hor}$.
Such a bound would enable one to rule out some of the proposed  models for the BH interior.

Here, we mainly focused on answering the binary question: Are the BHs in Nature  those  of  GR or are they not? If the latter is indeed true, further discrimination will be possible by looking at  other classes of fluid modes,
 as most of these  carry  information about the interior composition already at leading order in frequency. In these cases, however, the theoretical predictions will necessarily be model dependent. A detailed discussion of this topic from the perspective of the collapsed-polymer model \cite{strungout} can be found
in \cite{collision}. Other relevant works in this direction include \cite{Cardoso,Cardoso:2016oxy,noncardoso,Abedi:2017isz,Ashton:2016xff,barcelo}.

Finally, one might be concerned as to (i)  how interior fluid
modes can couple to external GWs in models with a  horizon, albeit a horizon with a quantum disposition, and (ii)
how an external observer would perceive this class of  GWs
in a way that is consistent with classical GR (which certainly maintains
its validity in the exterior part of spacetime).
As these are important issues in their own right, we intend to address
them in a separate discussion \cite{ridethewave}.

\subsection*{Acknowledgments}

We thank Paolo Pani for helpful comments on the damping time of a wormhole.
The research of RB was supported by the Israel Science Foundation grant no. 1294/16. The research of AJMM received support from an NRF Incentive Funding Grant 85353 and an  NRF Competitive Programme Grant 93595.  AJMM thanks Ben Gurion University for their  hospitality during his visit. KY acknowledges support from JSPS Postdoctoral Fellowships for Research Abroad, NSF grant PHY-1305682 and the Simons Foundation.

\end{document}